\documentclass[a4paper]{article}
\usepackage{INTERSPEECH2022}

\usepackage{amsmath,comment}
\usepackage{
amsfonts,
mathrsfs,
url,
xurl,
cite,
booktabs,
tabularx,
array,
multirow,
}

\newlength{\tablelength}
\newcommand{\myqt}[1]{``#1''}
\newcommand{\mysubsubsection}[1]{
    \subsubsection{#1}
    }
    
\newcommand{\mysubsection}[1]{
    \subsection{#1}
    }

\newcolumntype{C}{>{\centering\arraybackslash}X}
\newcolumntype{L}{>{\raggedleft\arraybackslash}X}

\makeatletter
\newcommand{\figcaption}[1]{\def\@captype{figure}\caption{#1}}
\newcommand{\tblcaption}[1]{\def\@captype{table}\caption{#1}}
\makeatother

\title{Introducing Auxiliary Text Query-modifier \\to Content-based Audio Retrieval}
\name{Daiki Takeuchi, Yasunori Ohishi, Daisuke Niizumi, Noboru Harada, and Kunio Kashino}
\address{
NTT Corporation, Japan
}
\email{daiki.takeuchi.ux@hco.ntt.co.jp}

\begin{document}

\maketitle

\begin{abstract} 
The amount of audio data available on public websites is growing rapidly, and an efficient mechanism for accessing the desired data is necessary.
We propose a content-based audio retrieval method that can retrieve a target audio that is similar to but slightly different from the query audio by introducing auxiliary textual information which describes the difference between the query and target audio.
While the range of conventional content-based audio retrieval is limited to audio that is similar to the query audio, the proposed method can adjust the retrieval range by adding an embedding of the auxiliary text query-modifier to the embedding of the query sample audio in a shared latent space.
To evaluate our method, we built a dataset comprising two different audio clips and the text that describes the difference.
The experimental results show that the proposed method retrieves the paired audio more accurately than the baseline.
We also confirmed based on visualization that the proposed method obtains the shared latent space in which the audio difference and the corresponding text are represented as similar embedding vectors.


\end{abstract}


\section{Introduction}
\label{sec:intro} 
A massive amount of audio data is available on public websites, and it will continue to increase.
Audio retrieval and environmental sound recognition by deep learning have been widely explored as way to use audio data effectively~\cite{gemmeke2017audio, fonseca2020fsd50k, kim2019audiocaps, drossos2019clotho, takeuchi2020effects, gong2021ast, mesaros2016tut, elizalde2019cross, Oncescu2021audio, Koepke2022audio, xie2022dcase, panyapanuwat2019unsupervised, ikawa2018acoustic, manocha2018content, kim2019improving, elizalde2022clap}.
Audio retrieval is particularly essential for extracting desired audio data from the massive amount of available data.

The audio retrieval methods using an audio query are called {\it content-based audio retrieval}. Those methods retrieve audio data with acoustic features similar to the query~\cite{wold1996content, sundaram2008audio, makinen2012evolutionary, manocha2018content, kim2019improving}.
The methods can retrieve any audio data as long as it is similar to the audio query.
However, it is necessary to prepare an audio query similar to the target audio data in advance.

On the other hand, the audio retrieval methods using a text query retrieve audio data paired with the text query ~\cite{kim2010using, elizalde2019cross, Oncescu2021audio, Koepke2022audio}.
Some methods retrieve audio clips associated with the text query, and others retrieve audio clips whose embedding in the latent space is similar to that of the text query.
Compared to content-based audio retrieval, we can easily prepare and edit text queries.
However, it is not always easy to describe the audio contents precisely in text.

%



To overcome the above issues, we propose a new content-based audio retrieval framework that combines the auxiliary text query modifier with the given audio query. 
With this approach, one possible scenario could be as follows:
First, we retrieve the initial audio by a method such as a text-query-based one, in which case the initial audio may not be close enough to the target one.
Next, we find another audio clip by using a query that combines the initial audio and an additional text that describes the difference between the target and the initial audio. This combines the content-based and the text-query-based methods. 
We expect that the second search result gets closer to the target audio; then, we can further repeat the second step until the result becomes acceptable.

\begin{figure}[t]
  \centering
\includegraphics[width=0.99\columnwidth]{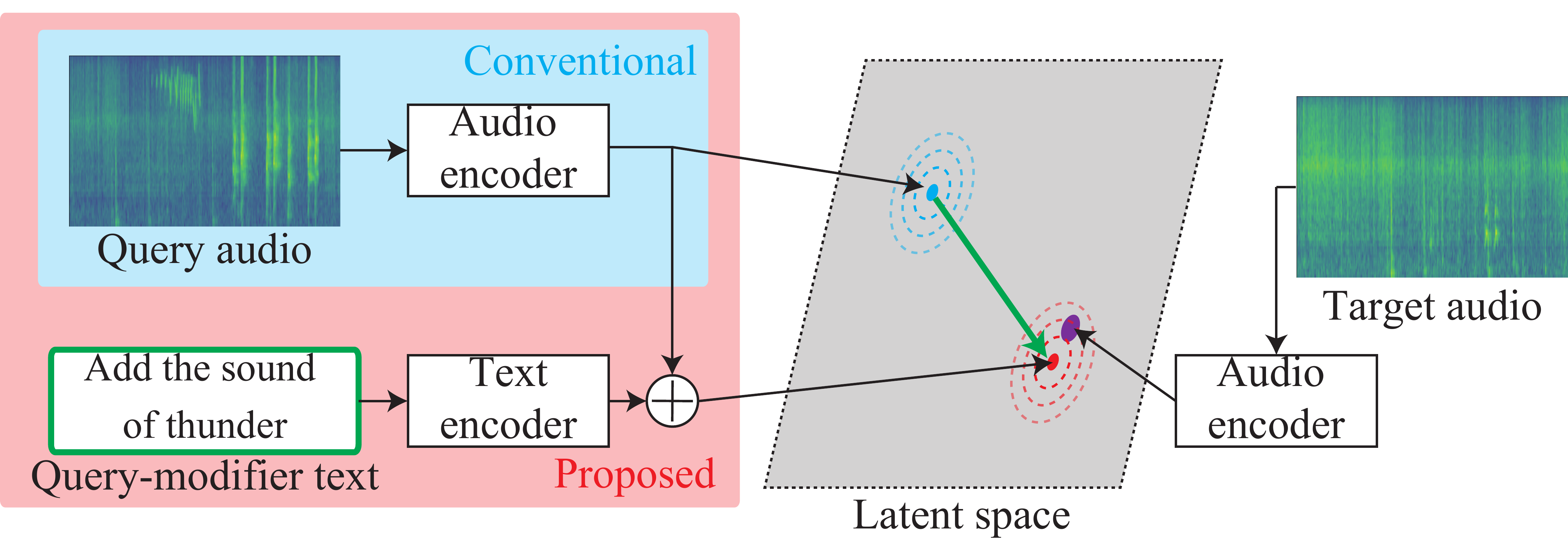} 
  \vspace{0pt}
  \caption{
  Conceptual diagram of proposed method. 
  In conventional content-based audio retrieval, a query audio becomes an embedding in a range (blue) in the latent space, roughly similar to the target audio (purple dot).
  The proposed method adds the embedding of the query-modifier text (green) to the query audio embedding, adjusting the retrieval range (red) closer to the target audio (purple dot).
  }
  \label{fig:overview}
  \vspace{-0pt}
\end{figure}

We implemented the framework as a method using neighborhood search in the latent space where the audio and difference described by text are embedded together.
The method learns the representations in the latent space using crossmodal contrastive learning.
We also introduced multi-task learning with a loss function that learns to classify audio contents to enhance differences in audio clips.
To evaluate the proposed method, we built a dataset comprising two sounds with a difference and a description of the difference.
Then, we conducted a comparison experiment with a baseline method that only uses sample audio as the query. We evaluated the retrieval accuracy and visualized the embeddings in the shared latent space and verified that the proposed method can adjust the retrieval range by means of the text query-modifier.

\section{Related work}
In content-based audio retrieval, audio data with features similar to the audio query in the latent space are retrieved.
Wold et al. attempted to retrieve audio on the basis of acoustic features, including loudness, pitch, brightness, and bandwidth~\cite{wold1996content}.
Guo et al. proposed using Mel-frequency cepstral coefficients as a feature and the latent space trained by a support vector machine\cite{guo2003content}.
Other methods used DNNs as embedding models to construct the latent space.
They trained the embedding models based on contrastive learning~\cite{manocha2018content} and unsupervised learning~\cite{panyapanuwat2019unsupervised}.
The advantage of the content-based method is that it can search for similar audio at the level of acoustic features, making it possible to retrieve audio closer to the query in terms of the acoustic feature representation.
On the other hand, the retrieval range is limited to audio similar to the query audio.
Since editing the speech waveform is a complex task, it is difficult to adjust the retrieval range by editing the query.

Audio retrieval methods using a text query train the common latent space using audio-text paired data annotated in advance and retrieve the audio data whose embedding on the trained latent space is similar to the embedding of the text query.
Audio tags written in text~\cite{elizalde2019cross} and text captions of audio data~\cite{Oncescu2021audio,Koepke2022audio,xie2022dcase} have been used as the queries in these methods.
Ikawa et al. proposed a method using onomatopoeia as a query~\cite {ikawa2018acoustic}. 
The range of audio retrieval with a text query can be easily adjusted by editing the text query.
However, collecting many pairs consisting of audio and text data for training is time-consuming, and words and phrases that are not included in the training data cannot be used to edit the text query meaningfully.

Our work is also related to crossmodal audio retrieval.
\cite{suris2018cross} proposed a method to retrieve audio data corresponding to a video query.
\cite{boggust2019grounding, Rouditchenko2021AVLnet} proposed methods to retrieve audio captions that explain the content of a video query.
Guzhov et al. proposed a method to obtain the latent space integrating three modalities: audio, visual, and text~\cite{guzhov2021audioclip}.
Unlike these methods, which use a query that consists of a single modality, we use a query that consists of two modalities: an audio sample and an auxiliary text.
We believe that the retrieval range can be adjusted by means of the auxiliary text.



\section{Proposed method}

\begin{figure}[t]
  \centering
\includegraphics[width=0.99\columnwidth]{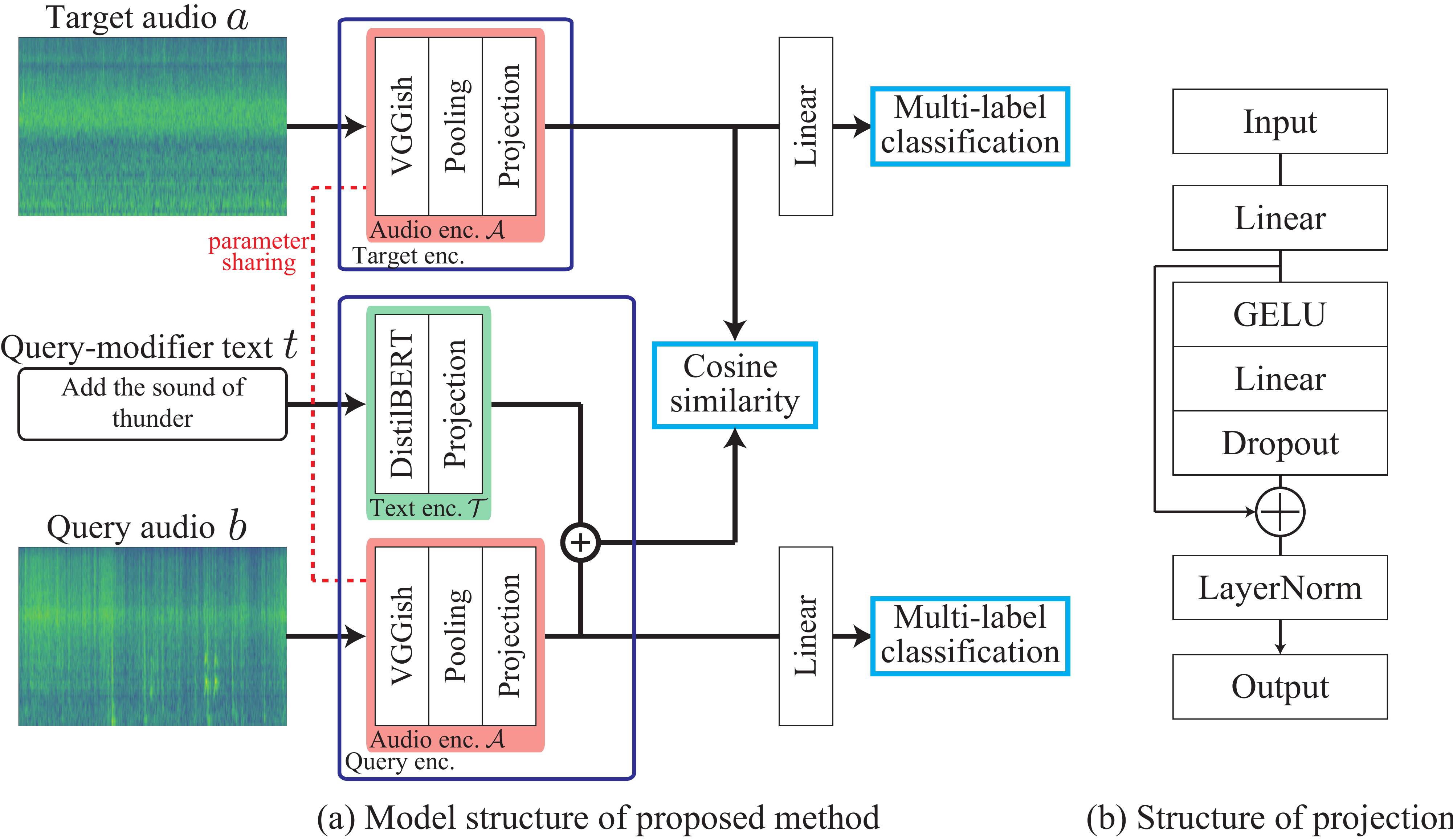} 
  \caption{Illustration of proposed method. The parameters of both audio encoders are shared. VGGish and DistilBERT pass the output of the last convolutional layer and the CLS token.}
  \label{fig:enc_structure}
\end{figure}

We propose a content-based audio retrieval method which can select the content of the retrieved audio data by adding a difference between the target audio data and the sample as auxiliary textual information when sample audio data close to the target one is given.
The proposed method uses sample audio and auxiliary text that describes the difference between the sample audio and the target audio as a search query.
We construct the method with a model structure and training method that can handle two modal queries.

Consider the problem of retrieving audio data $b$ from all candidate data by using audio query $a$ and text query $t$.
This can be regarded as a neighborhood search problem in the latent space, and can be treated as a problem of retrieving $b$ satisfying the following equation:
\begin{equation}
    \label{eq:base_problem}
    \min_{b}~d( \mathcal{F}_q(a, t),~ \mathcal{F}_t(b) ),
\end{equation}
where $\mathcal{F}_q$ is an embedding model for queries $a$ and $t$, $\mathcal{F}_t$ is an embedding model for target $b$, and $d$ is a distance function in the latent space.
The embedding model in Eq.~\eqref{eq:base_problem}, $\mathcal{F}_q$ and $ \mathcal{F}_t$, with an audio encoder and text encoder is as follows:
\begin{align}
    \label{eq:audio_text_enc}
    \mathcal{F}_q(a, t) = \mathcal{A}(a)+\mathcal{T}(t), \quad
    \mathcal{F}_t(b) = \mathcal{A}(b),
\end{align}
where $\mathcal{A}$ is an audio encoder, and $\mathcal{T}$ is a text encoder.
The parameters of the audio encoder $\mathcal{A}$ in $\mathcal{F}_q$ and $ \mathcal{F}_t$ are shared.
The structure of the audio encoder and text encoder is illustrated in Fig.~\ref{fig:enc_structure}(a).
In the audio encoder, the output from the last convolutional layer of the pre-trained VGGish~\cite{hershey2017cnn} is flattened in the channel direction, and the sum of its max pooling and mean pooling~\cite{kong2020panns} is input to the projection block.
In the text encoder, the text $t$ is input to the pretrained DistilBERT~\cite{sanh2019distilbert}, and the output corresponding to the CLS token, which is a special token prefixed with input text, is input to the projection block.
The projection block consists of two linear layers, the Gaussian error linear unit (GELU) function~\cite{hendrycks2016gaussian}, layer normalization, and a dropout layer as shown in  Fig.~\ref{fig:enc_structure}(b).
The cosine similarity is used for the distance function $d$.
Thus, Eq.~\eqref{eq:base_problem} is rewritten as
\begin{equation}
    \label{eq:problem_enc}
    \max_{b}~d_{\rm cossim}(\mathcal{A}(a)+\mathcal{T}(t), \mathcal{A}(b)),
\end{equation}
where $d_{\rm cossim}(x, y) = {xy}/{\|x\|_2\|y\|_2}$ is cosine similarity and $\|\cdot\|_2$ is $\ell^2$ norm.

In the proposed method, the audio encoder and the text encoder are trained by a crossmodal contrastive loss and an audio content classification loss.
The crossmodal contrastive loss connects the audio difference to its textual representation. 
The content classification loss trains the audio encoder so that the audio embedding has information about the type of content.
The use of this loss is intended to induce a clear representation of the type of difference in the content in the audio embedding differences.
Let us consider the set of data $\{a_n, b_n, t_n, v_n,  w_n\}_{n=1}^B$, 
where $a_n$ and $b_n$ are similar audio clips with some differences, 
$t_n$ is the description of the differences, 
$v_n$ and $w_n$ are the labels of the acoustic events contained in $a_n$ and $b_n$, 
$n$ is a data index, and $B$ is batch size.
The crossmodal contrastive loss function is the same as that used in \cite{radford2021learning} and is written as
\begin{equation}
    \label{eq:clip_loss}
    \mathcal{L}_{\rm cont} \!= \!
    - \frac{1}{2}\!\left(\!\sum_{i=1}^B \log\! \frac{ e^{z_{i,i}}} {\sum_{j=1}^B e^{z_{i,j}}}
    \!+\!\! \sum_{i=1}^B \log\! \frac{ e^{z_{i,i}}} {\sum_{j=1}^B e^{z_{j,i}}} \right)\!,
\end{equation}
where
\begin{equation}
    \label{eq:cossim_loss}
    z_{k,l} = d_{\rm cossim}(\mathcal{A}(a_k)+\mathcal{T}(t_k), \mathcal{A}(b_l))\cdot e^{\tau},
\end{equation}
$\tau$ is a temperature parameter which is fixed to $0$ for simplicity in the proposed method, and $i$ and $j$ are the indexes of data in a batch
The audio content classification loss is written as
\begin{equation}
    \label{eq:class_loss}
    \mathcal{L}_{\rm classif} \!=\!\!\sum_{i=1}^B \!\left\{ 
    \mbox{\footnotesize BCE}(\mathcal{C}(\mathcal{A}(a_i)),\! v_i)
    \!+\!
    \mbox{\footnotesize BCE}(\mathcal{C}(\mathcal{A}(b_i)),\! w_i) 
    \right\}\!,
\end{equation}
where  $\mbox{BCE}$ is the binary cross entropy loss and $\mathcal{C}$ is the linear layer with the sigmoid function.
Finally, the sum of the two losses is the overall loss for training: $\mathcal{L} = \mathcal{L}_{\rm cont}+\rho\mathcal{L}_{\rm classif}$, where $\rho$ is a weighting parameter.

In addition, the parameters of VGGish and DistilBERT are fixed in training.
Thus, only the parameters of the projection blocks are updated.

\begin{table}[!t]
\caption{Labels used to synthesize APwD-Dataset}
\vspace{-7pt}
\label{tab:label_list}
\centering
\scriptsize
\begin{tabularx}{0.95\columnwidth}{l|C|C }
\toprule
&\textit{Rain}&\textit{Traffic}\\
\midrule
background & \multirow{2}{*}{rain} & \multirow{2}{*}{car\_passing\_by}\\
(from FSD50K) && \\
\midrule
event &dog, chirping\_birds & dog, chirping\_birds \\
(from ESC-50)&thunder, footsteps &  car\_horn, church\_bells \\
\bottomrule
\end{tabularx}
\vspace{-10pt}
\end{table}

\section{Experiment}

In the experiment, we focused on three types of difference: an increase/decrease in a background sound, addition/removal of a sound event, and an increase/decrease of a sound event.
We built a dataset consisting of a pair of sounds with the above differences and the corresponding text and evaluated the proposed method by training and testing with its dataset.

\mysubsection{Audio Pair with Difference Dataset}
We built an Audio Pair with Difference Dataset~(APwD-Dataset) to evaluate the proposed method.
The APwD-Dataset was a set of two similar audio clips synthesized using the FSD50K~\cite{fonseca2020fsd50k} and ESC-50~\cite{piczak2015dataset} audio data, and an auxiliary text describing the differences between the similar audios.
Scaper~\cite{salamon2017scaper} was used to synthesize the similar audio data.
In this experiment, the dataset was created by setting up two scenes: \textit{Rain} and \textit{Traffic}~\footnote{
The dataset is available at https://github.com/nttcslab/apwd-dataset
}.
To synthesize a \textit{Rain} scene, data labeled \myqt{rain} in FSD50K was used as background, and data labeled \myqt{dog}, \myqt{chirping\_bird}, \myqt{thunder}, or \myqt{footsteps} in ESC-50 was used as events added to background.
To synthesize a \textit{Traffic} scene, data labeled \myqt{car\_passing} in FSD50K was used as background, and data labeled \myqt{dog}, \myqt{chirping\_birds}, \myqt{car\_horn}, or \myqt{church\_bells} in ESC-50 was used as events.
The sounds with specific labels used for synthesis in the FSD50K and ESC-50 shown in Table~\ref{tab:label_list}.
The development set and evaluation set for each scene contain 50,000 and 1,000 data, respectively.

The APwD dataset was synthesized using the following procedure.
First, the audio data with the labels were extracted from FSD50K and ESC-50.
The data assigned to the training and validation split of FSD50K and folds 1--4 of ESC-50 were used to synthesize the development set of the APwD-dataset, and the data assigned to the evaluation split of FSD50K and fold 5 of ESC-50 were used to synthesize the evaluation set.
Afterwards, audio data containing audible noise and other audio events were manually excluded.

\begin{figure*}[!t]
\vspace{-0pt}
\centering
\begin{tabular}{m{\textwidth}}
\hspace{0.05\columnwidth}
  \begin{minipage}{0.9\textwidth}
   \vspace{10pt}
  	\tblcaption{Comparison between proposed and baseline method}
  	\vspace{-5pt}
    \label{tab:result_retrieval}
    \footnotesize
    \scriptsize
    \small
    \begin{tabularx}{\columnwidth}{l|c|c| CCC | CCC }
    \toprule
    &&&\multicolumn{3}{c|}{\textbf{\textit{Rain}}}&\multicolumn{3}{c}{\textbf{\textit{Traffic}}} \\
    \textbf{Method} & \textbf{Text info.} & \textbf{Classif. loss} & \textbf{R@1} & \textbf{R@5} & \textbf{R@10} & \textbf{R@1} & \textbf{R@5} & \textbf{R@10} \\	
    \midrule
    (a) Baseline & $\times$ & $\times$ & 0.256 & 0.611 & 0.699 & 0.260 & 0.500 & 0.593 \\
    \midrule
    (b) Proposed w/o classif. loss & $\checkmark$ & $\times$ & 0.388 & 0.681 & 0.745 & 0.361 & 0.590 & 0.675 \\
    \midrule
    (c) Proposed w/ classif. loss & $\checkmark$ & $\checkmark$ & \textbf{0.445} & \textbf{0.721} & \textbf{0.769} & \textbf{0.391} & \textbf{0.622} & \textbf{0.695} \\
    \bottomrule
    \end{tabularx}
	
  \end{minipage}  \\
\hspace{0.05\columnwidth}
  \begin{minipage}{0.9\textwidth}
    \vspace{15pt}
    \tblcaption{R@1 comparison for different audio events}
    \vspace{-5pt}
    \label{tab:result_retrieval_each_diff}
    \scriptsize
    \begin{tabularx}{\columnwidth}{l|c|CCCCCCC}
    \toprule
    \textbf{Method} & \textbf{Scene}
    & $\textbf{background}^{\ast_1}$ & $\textbf{dog}^{\ast_2}$ & $\!\!\!\!\!\textbf{chirping\_bird}^{\ast_2}$ 
    & $\textbf{thunder}^{\ast_2}$ & $\textbf{footsteps}^{\ast_2}$ & $\textbf{car\_horn}^{\ast_2}$ &$\!\!\!\!\!\textbf{church\_bells}^{\ast_2}$\\
    \midrule
    (a) Baseline & \textit{Rain} 
    & 0.035 & 0.304 & 0.205 & 0.346 & 0.295 & N/A & N/A \\
    \midrule
    (b) Proposed w/o classif. loss & \textit{Rain} 
    & \textbf{0.061} & 0.448 & 0.360 & 0.490 & 0.450 & N/A & N/A \\
    \midrule
    (c) Proposed w/ classif. loss & \textit{Rain} 
    & \textbf{0.061} & \textbf{0.538} & \textbf{0.438} & \textbf{0.534} & \textbf{0.522} & N/A & N/A \\
    \midrule\midrule
    (a) Baseline  & \textit{Traffic} 
    & 0.056 & 0.327 & 0.164 & N/A & N/A & 0.369 & 0.294  \\
    \midrule
    (b) Proposed w/o classif. loss& \textit{Traffic}
    & 0.092 & 0.429 & 0.298 & N/A & N/A & 0.470 & 0.396 \\
    \midrule
    (c) Proposed w/ classif. losss & \textit{Traffic}
    &\textbf{ 0.115} & \textbf{0.502} & \textbf{0.327} & N/A & N/A & \textbf{0.495} & \textbf{0.435} \\
    \bottomrule
    \multicolumn{9}{l}{\vspace{-5pt}}\\
    \multicolumn{9}{l}{{\scriptsize$^{\ast_1}$The background audio is given a difference of increase or decrease. \myqt{bg} is \myqt{rain} in the {\it Rain} scene, and \myqt{car\_passing\_by} in the {\it Traffic} scene.}} \\
     \multicolumn{9}{l}{\vspace{-7pt}}\\
    \multicolumn{6}{l}{{\scriptsize$^{\ast_2}$The event audio is given a difference of an addition, deletion, increase, or decrease.}} \\
    \end{tabularx}
    \vspace{25pt}
  \end{minipage}
  \\
  \noindent
  \centering
  \hspace{-0.09\columnwidth}
  \begin{minipage}{0.97\textwidth}
    \begin{center}
    \centering
    \includegraphics[width=0.99\columnwidth]{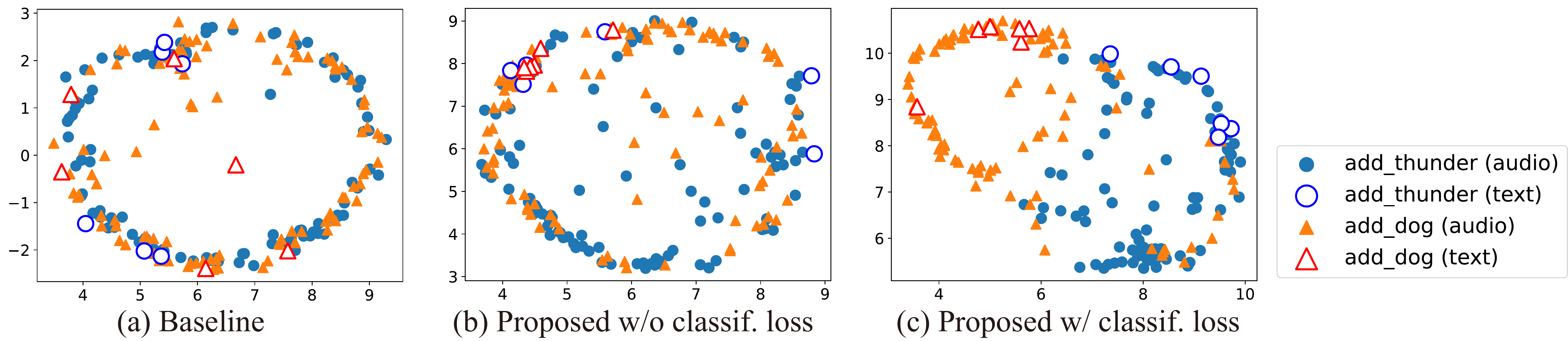} 
      \vspace{5pt}
      \figcaption{
      UMAP visualization of the difference in audio embedding vectors and text embedding vector.
      In (c), the proposed method with classification loss, the audio difference of ``the addition of thunder'' (blue circle) and that of ``the addition of dog'' (orange triangle) form different distributions for each type of difference.
      In addition, the text embedding~(thunder: blue circle outline, dog: red triangle outline) belongs to the same distribution as the corresponding difference in audio embedding vectors.
      }
      \label{fig:visualize}
      \vspace{0pt}
        \end{center}
  \end{minipage}
\end{tabular}
\end{figure*}

Next, we synthesized pairs of similar audio samples, $\alpha$ and $\beta$, which only consist of background audio.
Two pieces of data were cropped from the same audio file for 10 s at random locations and assigned to $\alpha$ and $\beta$.

Then, the difference was given between the pair of sound data by adding another background audio and/or event audio to $\alpha$ and $\beta$.
We consider the following six types of differences listed below and describe how they were synthesized.
Note that because it is difficult to semantically reduce or eliminate background audio and event audio, they are simulated by addition.
\begin{enumerate}
\def\labelenumi{(\alph{enumi})}
\setlength{\itemsep}{0pt}     
\setlength{\parskip}{0pt}      
\setlength{\itemindent}{0pt}   
\setlength{\leftskip}{-13pt}
\setlength{\labelsep}{1pt}     
\item Increase volume of background audio of $\alpha$: Randomly select the background sound data, cut out 10s, and add it to audio sample~$\beta$
\item Decrease volume of background audio of $\alpha$: Apply operation (a) and replace $\alpha$ and $\beta$
\item Add sound of audio event to $\alpha$: Randomly select the data used in the audio event and add it to audio sample~$\beta$
\item Remove sound of audio event from $\alpha$: Apply operation (c) and replace $\alpha$ and $\beta$
\item Increase volume of audio event of $\alpha$: Randomly select two audio data with the same label for the audio event and normalize them. After that, the amplitude of one data is reduced. The one with the larger amplitude is added to audio sample~$\beta$; the one with the smaller amplitude is added to audio sample~$\alpha$.
\item Decrease volume of audio event of $\alpha$: Apply operation (e) and replace $\alpha$ and $\beta$
\end{enumerate}
We provided one or two types of differences for each pair of audio samples. 

Finally, a description corresponding to the difference was assigned.
This description was written in the form of an imperative sentence, whose content described how to change the audio sample~$\alpha$ to the audio~$\beta$.
For example, if difference (a) is given to the paired data of the \textit{Rain} scene, the description is \myqt{increase the sound of rain}; if (f) is given for car\_horn and (c) for dog, the description is \myqt{make car horn lower and add dog bark.}

\mysubsection{Experimental conditions}
We used 10\% of the development set for validation.
The optimizer is Adam~\cite{kingma2014adam}.
The number of epochs was set to 300, and the model with the smallest loss of validation data was used for evaluation.
We used recall@$K$(R@$K$) to evaluate the accuracy of audio retrieval.
R@$K$ is the rate at which the ground-truth audio files are within the $K$th rank of the retrieval result.

We compared three models: the proposed method without classif. loss ($\rho\!\!=\!\!0$), the proposed method with classif. loss ($\rho\!\!=\!\!1$), and the baseline method.
For the baseline, we used the method of retrieving the target audio using only the sample audio without auxiliary text.
The parameters of the audio encoder were the same as in the proposed method.
Therefore, $z_{k,l} = d_{\rm cossim}(\mathcal{A}(a_k), \mathcal{A}(b_l))$
was applied to Eq.~\eqref{eq:clip_loss} instead of Eq.~\eqref{eq:cossim_loss}
In a preliminary study, we examined another method that takes an existing captioning system and a text query-modifier. However, it was not as effective as the baseline method using a sample audio query, so we did not include it in the comparison.

\mysubsection{Results}
\mysubsubsection{Comparison between proposed method and baseline}

We conducted an experiment to compare the proposed method with the baseline, the results of which are shown in~Table~\ref{tab:result_retrieval}.
Bold font indicates the highest scores.
The retrieval accuracy of the proposed method was higher than that of the baseline method in all conditions.
Table~\ref{tab:result_retrieval_each_diff} shows the Recall@1 for each background and event sound that was given a difference.
\myqt{background} is \myqt{rain} in the {\it Rain} scene and \myqt{car\_passing\_by} in the {\it Traffic} scene.
The results show that the proposed method outperformed the baseline method for all background and event sounds.
The retrieval accuracy of the differences in the background audio of each scene, \myqt{rain} and \myqt{car\_passing}, was significantly lower than the others.
Thus, dealing with differences in the background audio was more difficult than dealing with one in the audio event.


\mysubsubsection{UMAP visualization of embedding vector}
To verify that the proposed method handles difference information appropriately, we visualized the embedding vectors in the latent space.
For visualization, we created 100 datasets each for the difference between ``the addition of thunder'' and ``the addition of dog''.
The difference in the embedding vectors of audio data pairs, $\mathcal{A}(b)-\mathcal{A}(a)$, and the embedding vector of text, $\mathcal{T}(t)$, were visualized using UMAP~\cite{mcinnes2018umap}.

The visualization results are shown in Fig.~\ref{fig:visualize}, where (a), (b), and (c) show the embedding vectors for the baseline, the proposed method without classification loss, and the proposed method with classification loss, respectively.
In the baseline, embedding vectors were placed regardless of the type of difference.
In the proposed method without classification loss, the distributions of the embedding vectors of ``the addition of thunder'' and ``the addition of dog'' were slightly biased to the lower right and upper left, respectively.
In the proposed method with classification loss the distributions of the embedding vectors of ``the addition of thunder'' and ``the addition of dog'' were clearly biased to the lower right and upper left, respectively. 
Focusing on the text embedding vectors, 
they were placed such that they belonged to the same distribution as the corresponding difference of audio embedding vectors in the proposed method with classification loss.
Therefore, the training classification task for audio data should enhance the recognition of differences.




\section{Conclusion}
We proposed a content-based audio retrieval method that can retrieve target audio that is similar to but slightly different from the query audio by introducing an auxiliary text-query modifier which describes the difference between the query and the target audio.
The proposed method adjusts the retrieval range to obtain the target sound by adding the embedding vector of the auxiliary text-query modifier to that of the query audio in shared latent space.
We experimentally verified that the proposed method can obtain audio data with the difference from the query sound by utilizing the information in the introduced auxiliary text.
After visualizing the embedding vectors in the latent space, we also verified that the proposed method learns the relation between the audio difference and its textual representation.

Future work includes improving the accuracy of the search for increases and decreases in background so that it as accurate as the search for differences in audio events.
We also aim to collect data from actual recordings annotated by humans to build a model that can handle a wider variety of descriptions and differences.
%



\clearpage
\begin{thebibliography}{99}
  
  \bibitem{gemmeke2017audio}
  J.~F. Gemmeke, D.~P. Ellis, D.~Freedman, A.~Jansen, W.~Lawrence, R.~C. Moore,
    M.~Plakal, and M.~Ritter, ``Audio set: An ontology and human-labeled dataset
    for audio events,'' in \emph{Proc. IEEE Int. Conf. Acoust. Speech Signal
    Process. (ICASSP)}.\hskip 1em plus 0.5em minus 0.4em\relax IEEE, 2017, pp.
    776--780.
  
  \bibitem{fonseca2020fsd50k}
  E.~Fonseca, X.~Favory, J.~Pons, F.~Font, and X.~Serra, ``Fsd50k: an open
    dataset of human-labeled sound events,'' \emph{arXiv preprint
    arXiv:2010.00475}, 2020.
  
  \bibitem{kim2019audiocaps}
  C.~D. Kim, B.~Kim, H.~Lee, and G.~Kim, ``{AudioCaps}: Generating captions for
    audios in the wild,'' in \emph{Proc. Conf. N. Am. Chapter Assoc. Comput.
    Linguist.}, 2019, pp. 119--132.
  
  \bibitem{drossos2019clotho}
  K.~Drossos, S.~Adavanne, and T.~Virtanen, ``Clotho: An audio captioning
    dataset,'' in \emph{Proc. IEEE Int. Conf. Acoust. Speech Signal Process.
    (ICASSP)}, 2019, pp. 736--740.
  
  \bibitem{takeuchi2020effects}
  D.~Takeuchi, Y.~Koizumi, Y.~Ohishi, N.~Harada, and K.~Kashino, ``Effects of
    word-frequency based pre- and post- processings for audio captioning,'' in
    \emph{Proc. Detect. Classif. Acoust, Scenes Events Workshop (DCASE)},
    November 2020.
  
  \bibitem{gong2021ast}
  Y.~Gong, Y.-A. Chung, and J.~Glass, ``{AST}: Audio spectrogram transformer,''
    in \emph{Proc. Interspeech}, 2021, pp. 571--575.
  
  \bibitem{mesaros2016tut}
  A.~Mesaros, T.~Heittola, and T.~Virtanen, ``{TUT} database for acoustic scene
    classification and sound event detection,'' in \emph{Proc. 24th Eur. Signal
    Process. Conf. (EUSIPCO)}.\hskip 1em plus 0.5em minus 0.4em\relax IEEE, 2016,
    pp. 1128--1132.
  
  \bibitem{elizalde2019cross}
  B.~Elizalde, S.~Zarar, and B.~Raj, ``Cross modal audio search and retrieval
    with joint embeddings based on text and audio,'' in \emph{Proc. IEEE Int.
    Conf. Acoust. Speech Signal Process. (ICASSP)}.\hskip 1em plus 0.5em minus
    0.4em\relax IEEE, 2019, pp. 4095--4099.
  
  \bibitem{Oncescu2021audio}
  A.-M. Oncescu, A.~Koepke, J.~Henriques, Z.~Akata, and S.~Albanie, ``Audio
    retrieval with natural language queries,'' in \emph{Proc. Interspeech}, 2021.
  
  \bibitem{Koepke2022audio}
  A.~S. Koepke, A.-M. Oncescu, J.~Henriques, Z.~Akata, and S.~Albanie, ``Audio
    retrieval with natural language queries: A benchmark study,'' \emph{IEEE
    Trans. Multimedia}, 2022.
  
  \bibitem{xie2022dcase}
  H.~Xie, S.~Lipping, and T.~Virtanen, ``Dcase 2022 challenge task 6b:
    Language-based audio retrieval,'' \emph{arXiv preprint arXiv:2206.06108},
    2022.
  
  \bibitem{panyapanuwat2019unsupervised}
  P.~Panyapanuwat, S.~Kamonsantiroj, and L.~Pipanmaekaporn, ``Unsupervised
    learning hash for content-based audio retrieval using deep neural networks,''
    in \emph{Proc. 11th Int. Conf. Knowl. Smart Technol. (KST)}.\hskip 1em plus
    0.5em minus 0.4em\relax IEEE, 2019, pp. 99--104.
  
  \bibitem{ikawa2018acoustic}
  S.~Ikawa and K.~Kashino, ``Acoustic event search with an onomatopoeic query:
    measuring distance between onomatopoeic words and sounds,'' in \emph{Proc.
    Detect. Classif. Acoust. Scenes Events (DCASE) Workshop}, 2018, pp. 59--63.
  
  \bibitem{manocha2018content}
  P.~Manocha, R.~Badlani, A.~Kumar, A.~Shah, B.~Elizalde, and B.~Raj,
    ``Content-based representations of audio using siamese neural networks,'' in
    \emph{Proc. IEEE Int. Conf. Acoust. Speech Signal Process. (ICASSP)}.\hskip
    1em plus 0.5em minus 0.4em\relax IEEE, 2018, pp. 3136--3140.
  
  \bibitem{kim2019improving}
  B.~Kim and B.~Pardo, ``Improving content-based audio retrieval by vocal
    imitation feedback,'' in \emph{Proc. IEEE Int. Conf. Acoust. Speech Signal
    Process. (ICASSP)}.\hskip 1em plus 0.5em minus 0.4em\relax IEEE, 2019, pp.
    4100--4104.
  
  \bibitem{elizalde2022clap}
  B.~Elizalde, S.~Deshmukh, M.~A. Ismail, and H.~Wang, ``Clap: Learning audio
    concepts from natural language supervision,'' \emph{arXiv preprint
    arXiv:2206.04769}, 2022.
  
  \bibitem{wold1996content}
  E.~Wold, T.~Blum, D.~Keislar, and J.~Wheaten, ``Content-based classification,
    search, and retrieval of audio,'' \emph{IEEE multimedia}, vol.~3, no.~3, pp.
    27--36, 1996.
  
  \bibitem{sundaram2008audio}
  S.~Sundaram and S.~Narayanan, ``Audio retrieval by latent perceptual
    indexing,'' in \emph{Proc. IEEE Int. Conf. Acoust. Speech Signal Process.
    (ICASSP)}.\hskip 1em plus 0.5em minus 0.4em\relax IEEE, 2008, pp. 49--52.
  
  \bibitem{makinen2012evolutionary}
  T.~M{\"a}kinen, S.~Kiranyaz, J.~Raitoharju, and M.~Gabbouj, ``An evolutionary
    feature synthesis approach for content-based audio retrieval,'' \emph{EURASIP
    J. Audio Speech Music Process.}, vol. 2012, no.~1, pp. 1--23, 2012.
  
  \bibitem{kim2010using}
  S.~Kim, P.~Georgiou, S.~Narayanan, and S.~Sundaram, ``Using na{\"\i}ve text
    queries for robust audio information retrieval,'' in \emph{Proc. IEEE Int.
    Conf. Acoust. Speech Signal Process. (ICASSP)}.\hskip 1em plus 0.5em minus
    0.4em\relax IEEE, 2010, pp. 2406--2409.
  
  \bibitem{guo2003content}
  G.~Guo and S.~Z. Li, ``Content-based audio classification and retrieval by
    support vector machines,'' \emph{IEEE trans. Neural Netw.}, vol.~14, no.~1,
    pp. 209--215, 2003.
  
  \bibitem{suris2018cross}
  D.~Sur{\'\i}s, A.~Duarte, A.~Salvador, J.~Torres, and X.~Gir{\'o}-i Nieto,
    ``Cross-modal embeddings for video and audio retrieval,'' in \emph{Proc. Eur.
    Conf. Comput. Vis. (ECCV) Workshops}, 2018, pp. 711--716.
  
  \bibitem{boggust2019grounding}
  A.~W. Boggust, K.~Audhkhasi, D.~Joshi, D.~Harwath, S.~Thomas, R.~S. Feris,
    D.~Gutfreund, Y.~Zhang, A.~Torralba, M.~Picheny, and J.~Glass, ``Grounding
    spoken words in unlabeled video.'' in \emph{Proc. Conf. Comput. Vis. Pattern
    Recognit. Workshops}, 2019, pp. 29--32.
  
  \bibitem{Rouditchenko2021AVLnet}
  A.~Rouditchenko, A.~Boggust, D.~Harwath, B.~Chen, D.~Joshi, S.~Thomas,
    K.~Audhkhasi, H.~Kuehne, R.~Panda, R.~Feris, B.~Kingsbury, M.~Picheny,
    A.~Torralba, and J.~Glass, ``{AVL}net: Learning audio-visual language
    representations from instructional videos,'' in \emph{Proc. Interspeech},
    2021.
  
  \bibitem{guzhov2021audioclip}
  A.~Guzhov, F.~Raue, J.~Hees, and A.~Dengel, ``{AudioCLIP}: Extending clip to
    image, text and audio,'' \emph{arXiv preprint arXiv:2106.13043}, 2021.
  
  \bibitem{hershey2017cnn}
  S.~Hershey, S.~Chaudhuri, D.~P.~W. Ellis, J.~F. Gemmeke, A.~Jansen, R.~C.
    Moore, M.~Plakal, D.~Platt, R.~A. Saurous, B.~Seybold, M.~Slaney, R.~J.
    Weiss, and K.~Wilson, ``{CNN} architectures for large-scale audio
    classification,'' in \emph{Proc. IEEE Int. Conf. Acoust. Speech Signal
    Process. (ICASSP)}.\hskip 1em plus 0.5em minus 0.4em\relax IEEE, 2017, pp.
    131--135.
  
  \bibitem{kong2020panns}
  Q.~Kong, Y.~Cao, T.~Iqbal, Y.~Wang, W.~Wang, and M.~D. Plumbley, ``{PANNs}:
    Large-scale pretrained audio neural networks for audio pattern recognition,''
    \emph{IEEE/ACM Trans. Audio Speech Lang. Process.}, vol.~28, pp. 2880--2894,
    2020.
  
  \bibitem{sanh2019distilbert}
  V.~Sanh, L.~Debut, J.~Chaumond, and T.~Wolf, ``{DistilBERT, a distilled version
    of BERT: smaller, faster, cheaper and lighter},'' \emph{arXiv preprint
    arXiv:1910.01108}, 2019.
  
  \bibitem{hendrycks2016gaussian}
  D.~Hendrycks and K.~Gimpel, ``Gaussian error linear units {(GELUs)},''
    \emph{arXiv preprint arXiv:1606.08415}, 2016.
  
  \bibitem{radford2021learning}
  A.~Radford, J.~W. Kim, C.~Hallacy, A.~Ramesh, G.~Goh, S.~Agarwal, G.~Sastry,
    A.~Askell, P.~Mishkin, J.~Clark, G.~Krueger, and I.~Sutskever, ``Learning
    transferable visual models from natural language supervision,'' \emph{arXiv
    preprint arXiv:2103.00020}, 2021.
  
  \bibitem{piczak2015dataset}
  K.~J. Piczak, ``{ESC}: {Dataset} for {Environmental Sound Classification},'' in
    \emph{Proc. 23rd {Annual ACM Conf. {Multimedia}}}.\hskip 1em plus 0.5em minus
    0.4em\relax {ACM Press}, pp. 1015--1018.
  
  \bibitem{salamon2017scaper}
  J.~Salamon, D.~MacConnell, M.~Cartwright, P.~Li, and J.~P. Bello, ``Scaper: A
    library for soundscape synthesis and augmentation,'' in \emph{Proc. IEEE
    Workshop Appl. Signal Process. Audio Acoust. (WASPAA)}.\hskip 1em plus 0.5em
    minus 0.4em\relax IEEE, 2017, pp. 344--348.
  
  \bibitem{kingma2014adam}
  D.~P. Kingma and J.~Ba, ``Adam: A method for stochastic optimization,'' in
    \emph{Proc. in Int. Conf. Learn. Represent. (ICLR)}, 2014.
  
  \bibitem{mcinnes2018umap}
  L.~McInnes, J.~Healy, and J.~Melville, ``{UMAP}: Uniform manifold approximation
    and projection for dimension reduction,'' \emph{arXiv preprint
    arXiv:1802.03426}, 2018.
  
  \end{thebibliography}
\end{document}